\documentclass[12pt,a4paper]{article}
\usepackage{amsmath,amssymb,amsfonts,xspace,mathrsfs}

\usepackage{jheppub}

\numberwithin{equation}{section}
\def\Im{\,{\rm Im}\,}
\def\Re{\,{\rm Re}\,}

\def\({\left(}
\def\){\right)}
\def\[{\left[}
\def\]{\right]}
\def\haf{\textstyle{1\over 2}}
\def\hf{\frac{1}{2}}

\renewcommand{\d}{\mathrm{d}}
\newcommand{\de}{\mathrm{d}}

\newcommand{\I}{\mathrm{i}}

\newcommand{\p}{\partial}

\newcommand{\cB}{\mathcal{B}}

\newcommand{\cV}{\mathcal{V}}

\newcommand{\cC}{\mathcal{C}}

\newcommand{\cG}{\mathcal{G}}
\newcommand{\cK}{\mathcal{K}}

\newcommand{\cM}{\mathcal{M}}
\newcommand{\cW}{\mathcal{W}}
\newcommand{\cN}{\mathcal{N}}
\newcommand{\cE}{\mathcal{E}}

\newcommand{\cJ}{\mathcal{J}}

\DeclareSymbolFont{AMSa}{U}{msa}{m}{n}
\DeclareSymbolFont{AMSb}{U}{msb}{m}{n}
\DeclareMathSymbol{\fieldR}{\mathalpha}{AMSb}{"52}

\newcommand{\kahler}{{K\"ahler}\xspace}
\newcommand{\hk}{{hyperk\"ahler}\xspace}
\newcommand{\qk}{{quaternion-K\"ahler}\xspace}

\newcommand{\cO}{\mathcal{O}}

\newcommand{\cU}{\mathcal{U}}
\newcommand{\cA}{\mathcal{A}}

\newcommand{\pa}{\partial}
\newcommand{\nn}{\nonumber}

\newcommand{\IR}{\mathbb{R}}

\newcommand{\IZ}{\mathbb{Z}}

\newcommand{\IP}{\mathbb{P}}

\newcommand{\Tr}{\mbox{Tr}}

\newcommand{\tzeta}{{\tilde\zeta}}

\newcommand{\CP}{\mathbb{P}}

\newcommand{\beq}{\begin{eqnarray}}
\newcommand{\eeq}{\end{eqnarray}}
\def\be{\begin{equation}}
\def\ee{\end{equation}}
\def\ba{\begin{align}}
\def\ea{\end{align}}
\def\bse{\begin{subequations}}
\def\ese{\end{subequations}}
\newcommand{\bea}[2]{\be\label{#2}\begin{array}{#1}}
\newcommand{\eea}{\end{array}\ee}

\def\ba{\bar a}

\def\bs{\bar s}
\def\bz{\bar z}

\def\bZ{\bar Z}

\def\bX{\bar X}
\def\bF{\bar F}

\def\ba{\bar a}

\def\bt{\bar t}

\def\bw{\bar w}

\def\bz{\bar z}

\def\tc{\tilde c}

\def\CY{\mathfrak{Y}}
\def\CYm{\hat\CY}

\def\CY{\mathfrak{Y}}
\def\CYm{\mathfrak{\hat Y}}
\def\Kf{\mathfrak{S}}

\def\Ma{M}
\def\Mi{M}

\def\matf{\mathfrak{f}}

\title{$\cN=2$ Heterotic-Type II duality and bundle moduli}

\preprint{CERN-PH-TH-2014-084, L2C:14-036,
ZMP-HH/14-11}

\author[a,b]{Sergei Alexandrov,}
\author[c,d]{Jan Louis,}
\author[e,f,g]{Boris Pioline,}
\author[h,i]{Roberto Valandro}

\affiliation[a]{Universit\'e Montpellier 2, Laboratoire Charles Coulomb UMR 5221, F-34095,
Montpellier, France}
\affiliation[b]{CNRS, Laboratoire Charles Coulomb UMR 5221, F-34095,
Montpellier, France}
\affiliation[c]{Fachbereich Physik der Universit\"at Hamburg, Luruper Chaussee 149, 22761 Hamburg, Germany}
\affiliation[d]{Zentrum f\"ur Mathematische Physik, Universit\"at Hamburg, Bundesstrasse 55, 20146 Hamburg, Germany}
\affiliation[e]{CERN PH-TH,
Case C01600, CERN, CH-1211 Geneva 23, Switzerland}
\affiliation[f]{CNRS, UMR 7589, LPTHE, F-75005, Paris, France}
\affiliation[g]{Sorbonne Universit\'es, UPMC Univ. Paris 06, UMR 7589, LPTHE, F-75005, Paris, France}
\affiliation[h]{ICTP, Strada Costiera 11, Trieste 34014, Italy}
\affiliation[i]{INFN, Sezione di Trieste, Italy}
\emailAdd{salexand@univ-montp2.fr}
\emailAdd{jan.louis@desy.de}
\emailAdd{boris.pioline@cern.ch}
\emailAdd{rvalandr@ictp.it}

\abstract{Heterotic string compactifications on a $K3$ surface $\mathfrak{S}$ depend on a choice of hyperk\"ahler metric,
anti-self-dual gauge connection and Kalb-Ramond flux, parametrized by hypermultiplet scalars. The metric on hypermultiplet
moduli space is in principle computable within the $(0,2)$ superconformal field theory on the heterotic string worldsheet, although little is known about it in practice. Using duality with type II strings compactified on a Calabi-Yau threefold,  we predict the form of the quaternion-K\"ahler metric on hypermultiplet moduli space when $\mathfrak{S}$ is elliptically fibered, in the limit of a large fiber and even larger base.
The result is in general agreement with expectations from Kaluza-Klein reduction, in particular the metric has a two-stage fibration structure, where the $B$-field moduli are fibered over bundle and metric moduli, while bundle moduli are themselves fibered over metric moduli. A more precise match must await a detailed analysis of $R^2$-corrected ten-dimensional supergravity.
}

\begin{document}
\maketitle

\section{Introduction}

Nearly thirty years after their inception \cite{Gross:1984dd,Candelas:1985en},
heterotic strings compactified on Calabi-Yau threefolds
continue to be a framework of choice for constructing MSSM or GUT-like perturbative string
vacua. However, despite this phenomenological appeal,
a complete description of the low-energy effective theory of such compactifications is still
 missing so far. This is due in part to the difficulty in understanding  the  gauge bundle moduli space,
but also to the subtle nature of the $B$-field in heterotic string theory \cite{Witten:1999eg,Witten:1999fq}.
Indeed, anomaly cancellation requires that $B$ transforms non-trivially under
diffeomorphisms and gauge transformations, so that the gauge invariant field strength
(with $\alpha'=1$)
\be
\label{eqBH}
H=\de B+\frac{1}{4}\,(\omega_{G} - \omega_{L} )
\ee
involves a contribution from the Chern-Simons forms $\omega_{G}$ and $\omega_{L}$,
for the gauge group $G$ ($E_8\times E_8$ or $SO(32)$) and the Lorentz group $SO(1,9)$, respectively.
The equation of motion and Bianchi identity
\be
\label{bianchi}
\de \star_g H=0\, ,
\qquad
\de H =\frac{1}{4} \( \Tr F\wedge F - \Tr R\wedge R\)  ,
\ee
 imply that $B$ cannot be expanded on a basis of harmonic forms, except in the special
 case of the standard embedding, where the r.h.s. of the Bianchi identity vanishes point wise.
 Furthermore, supersymmetry requires additional higher-derivative couplings in
 the $D=10$ supergravity Lagrangian \cite{Bergshoeff:1989de},
 which greatly complicate  the Kaluza-Klein reduction.

Even in the simpler case
of heterotic strings compactified on $K3$, preserving $\cN=1$ supersymmetries in 6 dimensions
(or equivalently, for heterotic strings on $K3\times T^2$, preserving $\cN=2$ supersymmetries
in 4 dimensions), the knowledge of the hypermultiplet moduli space $\cM_{\rm H}$ describing
metrics $g$, gauge bundles $F$ and $B$-fields on $K3$ is incomplete. In principle it is entirely
determined within the $(0,4)$ worldsheet SCFT at tree-level, by the usual decoupling argument
between hypermultiplets and vector multiplets, which include the heterotic dilaton together
with the metric and gauge bundle moduli on $T^2$ \cite{deWit:1984px}. The metric on $\cM_{\rm H}$ has
however remained largely unknown, aside  from the `standard embedding' locus
where  the SCFT has enhanced $(4,4)$ supersymmetry.
This ignorance, along with an
incomplete understanding of non-perturbative effects in type II Calabi-Yau vacua,
has prevented detailed tests of heterotic/type II duality \cite{Kachru:1995wm,Ferrara:1995yx} in the
hypermultiplet sector, beyond the early attempts in \cite{Aspinwall:1998bw,Aspinwall:1999xs,Aspinwall:2000fd}.
The situation on the type II side has considerably improved in recent years
(see \cite{Alexandrov:2011va,Alexandrov:2013yva} for recent reviews), and it is therefore natural
to strive for a similar improvement on the heterotic side. Although $\cN=2$ vacua are not phenomenologically relevant,
the lessons learned in this process will likely be useful for heterotic compactifications
on Calabi-Yau threefolds as well \cite{McOrist-in-progress}.

The Kaluza-Klein reduction of the ten-dimensional heterotic supergravity on $K3$ including bundle
and $B$-field moduli  was discussed recently in \cite{Louis:2011hp}, building on earlier work
\cite{Louis:2001uy,Louis:2009dq}. While the contribution of the gauge Chern-Simons form
$\omega_{G}$ to the equation of motion was included, the reduction did not include
the Lorentz Chern-Simons form $\omega_{L}$, nor the $R^2$ couplings related to it by supersymmetry.
As a result, the low-energy  effective action was not supersymmetric, in particular the metric on
hypermultiplet moduli space was not \qk (QK). One of the general lessons from
\cite{Louis:2011hp}, however, was that, unlike what has been often stated or implicitly assumed in the literature,
the hypermultiplet moduli space $\cM_{\rm H}$ is definitely {\it not}
a bundle of \hk (HK) spaces $\cM_{F}(g,V)$, parametrizing the bundle, over the moduli space
of the (4,4) SCFT, corresponding to the HK metric $g$ on $K3$ along with the $B$-field moduli.
Rather, it must have a two-stage fibration structure
\be
\label{doublefib}
\begin{array}{rccc}
  \cM_{B}(g,F) &\rightarrow & \cM_{\rm H} &\\
& & \downarrow &\\
\cM_{F}(g)& \rightarrow & \cM_{g,F}  & \\
& & \downarrow &\\
&& \cM_{g} &
\end{array}
\ee
where the $B$-field moduli $\cM_{B}(g,F)$ live in a torus bundle, fibered over both the metric moduli $\cM_g$
and the bundle moduli $\cM_F(g)$, the latter being fibered over $\cM_g$ as well. As
we discuss in \S\ref{sec_gen}, this structure is
an immediate consequence of the Bianchi identity \eqref{bianchi}, as recognized
early on in \cite{Witten:1999fq}. The main goal of this paper is to investigate
the structure of this two-stage
fibration and to understand how it can be compatible with the \qk (QK) property of the total
hypermultiplet moduli space $\cM_{\rm H}$, which is a necessary requirement for
supersymmetry \cite{Bagger:1983tt}.

For this purpose, our strategy will be to assume that heterotic/type II duality holds, use our
knowledge of the hypermultiplet moduli space on the type IIB side in a suitable limit,
and find a change of coordinates (or duality map) which displays the two-stage
fibration structure \eqref{doublefib}. In order for heterotic/type II duality to apply,
we assume that the $K3$ manifold $\Kf$ on the heterotic side is elliptically fibered,
while the Calabi-Yau three-fold on the type IIB side is
$K3$-fibered \cite{Louis:2011aa}.
Furthermore, we are interested in the limit where one can ignore $g_s$-corrections on the type IIB side
and $\alpha'$-corrections on the heterotic side.
To this end, we assume that the type IIB ten-dimensional string coupling is weak
(so that, on the heterotic side, the area of the elliptic fiber is much smaller than the base), and
that the volume of $\Kf$ is large in heterotic string units (so that, on the type IIB side,  the area of the base
of the $K3$ fibration is also large in type II string units). In this regime, and in the special
case of heterotic compactifications with a rigid bundle, a suitable duality map was constructed
in \cite{Louis:2011aa,Alexandrov:2012pr}, which identifies the hypermultiplet moduli space
on the type IIB side (parametrizing  the string coupling, Neveu-Schwarz axion, \kahler structures
and associated RR-potentials) with the heterotic hypermultiplet moduli space.

In this paper, we extend this duality map to the case of heterotic $K3$ compactifications
with non-rigid bundles. As in \cite{Louis:2011aa}, we identify the heterotic bundle moduli with the \kahler moduli
associated to reducible singular fibers on the type IIB side, along with the corresponding 
RR moduli.\footnote{This identification is analogous to the standard identification
between bundle moduli on $T^2$ and complex structure moduli associated to singular $K3$ fibers
of the CY 3-fold on the type IIA side \cite{Aspinwall:1995vk,Aspinwall:2000fd}.}
Retaining the first subleading correction to the holomorphic prepotential in the limit where the area of the
base of the $K3$ fibration is large, see \eqref{FclasG}, we find that the type IIB hypermultiplet space,
translated into heterotic variables, neatly displays the double fibration structure \eqref{doublefib}.
In particular, the HK metric on the bundle moduli space $\cM_F(g)$  is obtained by the rigid $c$-map
construction from the prepotential $f(t^i, t^\alpha)$ appearing as the subleading correction
in \eqref{FclasG}. This is partially expected since for elliptic $K3$'s the bundle moduli space
has the structure of a complex integrable system  \cite{Bershadsky:1997zs},
with a semi-flat metric in the limit where the area of the elliptic fiber is much smaller
than the base. It is however worth noting that the metric on $\cM_F(g)$ could have been
in the more general class of \hk metrics on cotangent bundles of \kahler manifolds constructed in \cite{zbMATH01582122}.
Another feature of the metric on $\cM_{\rm H}$ which can be extracted from the dual type II
description is the topology of the torus fiber $ \cM_{B}(g,F)$ over bundle moduli space.
We find that the curvature of the Levi-Civita connection on this torus bundle is in nice
agreement with the Bianchi identity \eqref{bianchi}. In addition, the classical hypermultiplet
metric predicts a precise fibration structure over metric moduli and 
various volume-suppressed corrections to the moduli space metric,
which are necessary for the \qk property. We leave it as  an open problem to derive these corrections by reducing
10D heterotic supergravity (including the higher order derivative corrections required
by supersymmetry \cite{Bergshoeff:1989de}) on $\Kf$.

The outline of this work is as follows. In \S\ref{sec_gen}, we discuss the general structure of
the hypermultiplet moduli space in heterotic strings compactified on elliptic $K3$ with a non-rigid bundle.
In \S\ref{sec_hetcmap}, we recall basic aspects of heterotic/type II duality and translate the classical
type IIB hypermultiplet metric in heterotic variables. We read off the two-stage
fibration \eqref{doublefib} and compare with expectations from the Kaluza-Klein reduction
of tree-level supergravity on the heterotic side. We conclude in \S\ref{sec_discuss}
with a discussion of open issues.

\section{Generalities on heterotic moduli spaces \label{sec_gen}}

In this section, we discuss qualitative aspects of the hypermultiplet moduli space in
compactifications of the heterotic string
on a $K3$ surface $\Kf$. The same hypermultiplet moduli space appears in compactifications
on $K3\times T^2$ down to 4 dimensions, as the additional scalar fields coming from the metric and
gauge bundle on  $T^2$ all lie in vector multiplets. For concreteness, we restrict our analysis to
elliptically fibered $K3$ surfaces and the gauge group $G=E_8\times E_8$, so that heterotic/type II duality applies,
but most of the considerations below hold more generally. Our notations
follow \cite{Alexandrov:2012pr}.

As mentioned in the introduction, vacua with unbroken $\cN=2$ supersymmetry are characterized by a \hk metric $g$ on
$\Kf$, a bundle $F$ on $\Kf$ with
second Chern class $c_2(F)=\chi(\Kf)=24$ (as follows from the Bianchi identity
\eqref{bianchi}) equipped with an anti-self dual connection $A$ such
that\footnote{We abuse notation and denote by the same letter $F$ the bundle and its curvature.}
$F= \de A+A\wedge A$, and a two-form $B$ on $\Kf$
satisfying the equation of motion \eqref{bianchi}.
The metric on the resulting moduli space turns out to be given by a sum of three terms corresponding
to the three types of the moduli (see \eqref{fullmetmat} below). In the following we discuss
each of these contributions separately.

\subsection{Metric moduli \label{secmetmod}}

For what concerns the metric degrees of freedom, it is well known that the moduli space of
smooth HK metrics on $\Kf$ is (up to global identifications) the homogeneous space
\be
\cM_{g} = \IR^+_{\rho} \times
\[\frac{SO(3,n-1)}{SO(3)\times SO(n-1)}\]_{\gamma_I^x}
\label{manifoldmet}
\ee
with $n=20$ (see e.g. \cite{Aspinwall:1996mn}). The first factor corresponds to the volume $\cV=e^{-\rho}$
in heterotic string units, whereas the second factor
is parametrized by periods of the triplet of HK forms $\cJ^x$, $x=1,2,3$,
 along a basis $\tau_I$ ($I=1,\dots, n+2$) of $H_2(\Kf,\IZ)$,
\be
\gamma_I^x = e^{\rho/2}\int_{\tau_I} \cJ^x\, ,
\qquad
\eta^{IJ} \gamma_I^x \gamma_J^y = 2 \delta^{xy}\, ,
\label{Jform}
\ee
where $\eta^{IJ}$ is the inverse of the intersection matrix on $H_2(\Kf,\IZ)$. The periods
$\gamma_I^x$ may be organized into a $SO(3,n-1)$ symmetric matrix
\be
\label{defMIJ}
\Mi_{IJ}=\int_{\Kf} \omega_I\wedge \star\omega_J
=-\eta_{IJ}+\gamma^x_I\gamma^x_J,
\ee
where $\omega_I$ is a 2-form dual to $\tau_I$,
satisfying the following property
\be
(\Mi^{-1})^{IJ}= \eta^{IK} \eta^{JL} \Mi_{KL}\equiv \Mi^{IJ}  \, .
\ee
The matrix $\Mi_{IJ}$ parametrizes the conformal class of the HK metric on $\Kf$.
In terms of this matrix and the volume coordinate $\rho$,  the $SO(3,n-1)$ invariant  metric
on $\cM_g$ reads
\be
\label{dsginv}
\de s^2_g =
\hf\, \de \rho^2  -\frac14\, \de M_{IJ}\de M^{IJ}\, .
\ee
For singular $K3$'s with
$k$ shrinking cycles the moduli space has the same structure as above, with $\tau_I$
running over a basis of $n+2=22-k$ unobstructed cycles.

For elliptically fibered $K3$, the description given above can be further refined.
Let us denote by $\cB$ the base of the elliptic fibration, $\cE$ the elliptic fiber,
and $\tau_A$ a basis of the transcendental lattice on $\Kf$.
Here the index $A$ is running over a set of cardinality $n$, and we choose the basis of $H_2(\Kf,\IZ)$
to be given by $\tau_I=(\cB+\cE,\cE,\tau_A)$ with $\{I\}=\{1,2,A\}$.
Furthermore, we can fix the $SO(3)$ rotation symmetry among the three complex structures
by choosing a complex structure  adapted to
the elliptic fibration, i.e. $\int_{\cE} \cJ=0$ where $\cJ=\cJ^1+\I\cJ^2$ 
is the holomorphic 2-form in this complex structure.
In this setup the intersection form $\eta_{IJ}$ is given by
\be
\eta_{IJ}=\( \begin{array}{ccc}
0 & 1 & 0
\\
1 & 0 & 0
\\
0 & 0 & \eta_{AB}
\end{array}\),
\label{interform}
\ee
where $\eta_{AB}$ is the intersection form on the transcendental lattice, 
and the periods $\gamma^x_I$ can be parametrized as
\be
\begin{split}
&
\gamma_1=-\eta^{AB}\gamma_A v_B,
\qquad\qquad\
\gamma_2=0,
\qquad\quad\
\gamma_A=\frac{X_A}{\sqrt{\eta^{BC}X_B\bX_C}},
\\
&
\gamma_1^3=e^{-R/2} -\frac{ v^2}{2}\, e^{R/2},
\qquad
\gamma_2^3=e^{R/2},
\qquad
\gamma_A^3=e^{R/2} v_A
\end{split}
\label{param-zetatree}
\ee
with $\gamma_I=\hf(\gamma_I^1+\I\gamma_I^2)$.
The complex structure moduli $X^A$ are subject to the constraint $X^A \eta_{AB} X^B=0$
following from the orthogonality relation \eqref{Jform}, while the $v_A$'s are the (real)
\kahler moduli. The complex structure moduli are conveniently
organized into a $SO(2,n-2)$ symmetric matrix
\be
\Ma_{AB}=-\eta_{AB}+2X_{AB}\, ,
\qquad
X_{AB}=\frac{X_A\bX_B+\bX_A X_B}{\eta^{CD}X_C\bX_D}\, ,
\label{matrixMAB}
\ee
which allows to rewrite the metric \eqref{dsginv} on $\cM_g$ as follows
\be
\label{dsg}
\de s^2_g =  \hf\, \de \rho^2+
\hf\, \de R^2 -\frac14\, \de M_{AB}\de M^{AB}+e^R M_{AB} \de v^A \de v^B \, ,
\ee
where the third term may be equivalently written as
\be
-\frac{1}{4}\, \de M_{AB}\de M^{AB}=4 \,\frac{\de X^A\de \bX^B}{\eta_{CD}X^C \bX^D}\(X_{AB}-\eta_{AB}\).
\ee
This implies that the second factor in \eqref{manifoldmet} has the following bundle structure
\be
\begin{array}{crc}
\displaystyle{\IR^{2,n-2}_{v^A}} & \longrightarrow &
\left[\displaystyle{\frac{SO(3,n-1)}{SO(3)\times SO(n-1)}}\right]_{\gamma_I^x}
\\
&& \downarrow
\\
& \displaystyle{\IR^+_R} & \displaystyle{\times
 \[\frac{SO(2,n-2)}{SO(2)\times SO(n-2)}\]_{X^A}}\, ,
\end{array}
\ee
where the fiber parametrizes the \kahler structure while the base parametrizes
the complex structure compatible with the elliptic fibration.

\subsection{Bundle moduli}

We now turn to the  gauge bundle moduli. For simplicity, we restrict to bundles whose
structure group lies in a subgroup $H=SU(N)\subset G$. This breaks the  gauge symmetry
 from $G$ down to $G_0$, where $G_0$ is the commutant\footnote{At special points in moduli space, e.g. when the bundle becomes point-like,
the gauge symmetry may enhance and some additional charged matter may appear. This symmetry enhancement is most easily understood in $D=4$,
where it corresponds to standard unHiggsing. Going to the Coulomb branch
of this gauge theory typically leads to a new family of vacua with different bundle topology.
In six dimensions the transition is more exotic and involves
tensionless strings \cite{Witten:1995gx,Seiberg:1996vs}.
In this paper we stay away from such transitions. }
of $H$ inside $G$.
The Bianchi identity \eqref{bianchi} fixes the second Chern class to be $c_2(F)=24$, while $c_1(F)=0$ for $SU(N)$ bundles.
Supersymmetry
requires the connection $A$ to be anti-self-dual, $\star F=-F$.

For fixed metric $g$ on $\Kf$
and bundle topology, the space of gauge inequivalent anti-self-dual connections is
a finite-dimensional \hk space $\cM_F(g)$ of quaternionic dimension \cite{zbMATH03900915}
\be
\label{mdim1}
m = c_2(F)\, h(H)-{\rm dim}(H)\, ,
\ee
where $h(H)$ is the dual Coxeter number of $H$, equal to $N$ for $H=SU(N)$.
The $S^2$ family of
complex structures on $\cM_F(g)$ arises from the fact that the anti-self dual
condition is equivalent to the hermitian-Yang-Mills equations  in any complex structure $J$
on $\Kf$ \cite{MR506229},
\be\label{HYMeq}
F^{2,0}=0\, ,
\qquad F \wedge \cJ^3=0\, .
\ee
Solutions to \eqref{HYMeq} are in one-to-one correspondence with
semi-stable holomorphic vector bundles \cite{zbMATH03840763},
thereby providing a complex structure on $\cM_F(g)$. The tangent space over a given connection
$A_0$ is generated  by hermitian one-forms  $\hat{a}\in \Omega^{1}(\mbox{End}^hF)$ such that
\be
\de_{A_0} \hat{a} \in \Omega^{1,1}(\mbox{End}F)\, ,
\qquad
\int \cJ^3\wedge  \de_{A_0}\hat{a}=0\, ,
\qquad
\de_{A_0}^\dagger  \hat{a} =0\, ,
\ee
where
$\de_{A_0}\hat{a}=\de\hat{a}+[A_0,\hat a]$
and the last condition fixes the gauge. By using the complex structure, we can split the
connection $A_0$ and its variation $\hat{a}$ into
their (0,1) and (1,0) parts,
\begin{equation}
\label{eqtana0}
A_0=\cA_0+\cA_0^\dagger\, ,
\qquad
 \hat{a} = a + a^\dagger\, ,
 \qquad \mbox{where} \quad \cA_0,a \in \Omega^{0,1}(\mbox{End}F) \:.
\end{equation}
so that the tangent space over $A_0$ is isomorphic to the space of $\bar{\partial}_{\cA_0}$-harmonic (0,1)-forms
with values in End$F$, i.e. the cohomology group $H^{0,1}($End$F)$.
Hence, its complex dimension  $2m\equiv h^{0,1}($End$F)$ is equal to minus the index
\begin{equation}
 \chi(\mbox{End}F) = \int_\Kf \mbox{Td}(\Kf) \wedge {\rm ch}(\mbox{End}F)
 = 2 \, \mbox{rk}(\mbox{End}F) + {\rm ch}_2(\mbox{End}F) \:,
\end{equation}
where $\mbox{Td}(\Kf)=1+\tfrac{1}{12} c_2(\Kf)$ and ${\rm ch}={\rm rk}+c_1+\frac12 c_1^2-c_2$.
For an $SU(N)$ bundle,
${\rm rk}(\mbox{End}F) =N^2-1$, $c_1(\mbox{End}F) =0$, $c_2(\mbox{End}F) =2N c_2(F)$, in agreement with \eqref{mdim1}.
It is important to note that  the metric on $\cM_F(g)$
depends on the metric $g$ on $\Kf$ via the connection $A_0$, which is assumed to be anti-self-dual with respect to $g$.

We denote by $\xi^k$ ($k=1,\dots, 2m$) a system of complex coordinates on $\cM_{F}(g)$,
and choose a family  of solutions $\cA(y,\xi^k,g^i)$ of the anti-self-duality
equations which intersects each gauge orbit once (here $y$ denotes a set
of coordinates on $\Kf$, and $g^i$ denote the metric moduli discussed in \S\ref{secmetmod}).
Of course, $\cA(y,\xi^k,g^i)$ is ambiguous modulo gauge transformations
$\cA\to \cU\cA\cU^{-1}-\bar\pa \cU\, \cU^{-1}$.
In order to cancel this ambiguity,
one introduces
an ${\rm End}F$-valued connection $\Lambda=\Lambda_k \de\xi^k$ on $\cM_{F}(g)$
transforming as $\Lambda_k\to \cU\,\Lambda_k\,\cU^{-1}-\pa_{\xi^k} \cU\, \cU^{-1}$
in such a way that $D_{\xi^k} \cA\equiv \pa_{\xi^k}\cA-\bar\partial \Lambda_k+[\cA,\Lambda_k]$ transforms linearly,
$D_{\xi^k} \cA\to \cU \, D_{\xi^k}\cA \,\cU^{-1}$  \cite{McOrist-in-progress}.
By construction, $D_{\xi^k}\cA$ furnish a basis for $H^{0,1}($End$F)$.
Similarly, we need covariant derivatives
$D_{g^i} \cA$ with respect to metric moduli, which cancel the gauge dependence
of ordinary derivatives $\pa_{g^i} \cA$ (note that $D_{g^i}\cA$ are in general {\it not}
(0,1) forms, since a variation of $g^i$ may change the complex structure).
The fluctuation $a$ in \eqref{eqtana0} is then a linear combination
\be
\label{defa}
a = D_{\xi^k}\cA\, \de\xi^k+D_{g^i} \cA \, \de g^i
\ee
of the covariant derivatives of $\cA$.

In order to compute the metric on the bundle moduli space $\cM_F(g)$, one should  dimensionally reduce the gauge kinetic term
in $D=10$  heterotic supergravity, assuming that the gauge fields depend  
on non-compact directions $x$ through the bundle and metric moduli, $\xi^k(x)$ and $g^i(x)$.
The fluctuation \eqref{defa} then represents the part of the gauge field strength with
one leg on $\Kf$ and one leg in the non-compact directions.  This results in \cite{McOrist-in-progress}
\be
e^{\rho/2}\de s^2_{F} = e^{\rho} \, \cG_{k{\bar \ell}} (\de \xi^k+ C^k_i \de g^i)\,
(\de\bar \xi^{\bar \ell} + \bar{C}^{\bar \ell}_j \de g^j)\, ,
\label{metric-bundle}
\ee
where
\begin{eqnarray}
\cG_{k\bar \ell} &=& \int_\Kf \Tr \, D_{\xi^k}\cA \wedge \star D_{\bar{\xi}^\ell}\cA^\dagger\, ,
\qquad 
C^k_i = \cG^{k\bar{\ell}}  \int_\Kf  \Tr \, D_{\bar{\xi}^\ell}\cA^\dagger \wedge \star D_{g^i}\cA  \, .
\end{eqnarray}
By dimensional analysis, $\cG_{k\bar \ell}$ scales with the square root of the volume of $\Kf$,
so that \eqref{metric-bundle} scales as $e^{\rho/2}$, justifying the notation on the l.h.s.
For fixed geometric moduli, the metric $\cG_{k\bar \ell}$ is \kahler,
with \kahler potential $\int_\Kf \Tr \,  \cA\wedge \star \cA^\dagger$ \cite{zbMATH00044936}.

The previous discussion only relied on the Hermitian Yang-Mills equations \eqref{HYMeq}
in a fixed complex structure, and produced a \kahler metric on bundle moduli space $\cM_F(g)$
for any Calabi-Yau compactifications. For compactifications on a $K3$ surface $\Kf$,
the bundle moduli space $\cM_F(g)$ carries additional structure.  First, as already mentioned,
the equivalence with \eqref{HYMeq}
holds for any complex structure on $\Kf$, and produces a $S^2$ worth of complex structures on
$\cM_F(g)$. In addition, $\cM_F(g)$ inherits a holomorphic symplectic structure
from the holomorphic symplectic structure on $\Kf$ \cite{zbMATH03971677}.
The holomorphic
symplectic form on $\cM_F$ is given by the natural inner product on (End$F$-valued) one-forms,
\begin{equation}
\label{holsymF}
\langle\beta_1,\beta_2\rangle=\int_\Kf \Tr \, \beta_1\wedge \star \beta_2\:.
\end{equation}
Along with the
\kahler form,  the holomorphic symplectic form \eqref{holsymF} provides a \hk structure on $\cM_F(g)$.

In addition, when $\Kf$ is an elliptic fibration $\cE \to \Kf \to \cB$, $\cM_F(g)$ is in fact a complex
integrable
system \cite{Bershadsky:1997zs}. This can be seen from
spectral cover construction of bundles on elliptic
fibrations \cite{Friedman:1997yq,Donagi:1998vx} (see e.g.
\cite{Donagi:1999gc} for a non-technical discussion). The restriction of the $SU(N)$
bundle on the elliptic fibers produces an $N:1$ covering $\cC$ of the base $\cB$
known as the spectral (or cameral) curve. Using the isomorphism between
$\cE$ and its Jacobian, $\cC$ can be viewed as an effective curve inside $\Kf$ homologous
to $k \cE + N \cB$, where $k=c_2(F)$ \cite{Bershadsky:1997zs}. To recover the full bundle
on $\Kf$, it is necessary
to specify a degree $g-1-N$ line bundle on $\cC$, where $g=N(k-N)+1$ is the genus
of the curve $\cC$ \cite{Aspinwall:1998he}. The choice of spectral cover is parametrized by
a complex projective space  $\IP^{g}$ (the linear system associated to the divisor $k \cE + N \cB$),
while the choice of line bundle is parametrized by the
Jacobian torus of $\cC$, of complex dimension $g$. This realizes the complex integrable
system mentioned earlier. In practice, one can decompose the complex coordinates $\xi^k$
into (complex) action and angle variables $(t^\alpha,w_\alpha)$ such
that the holomorphic symplectic form is $\de t^\alpha \wedge \de w_{\alpha}$, while
the metric (at fixed metric moduli) becomes  
\be
\label{dsFcover}
e^{\rho/2}\de s^2_{F} = e^{\tfrac12(\rho-R)} \gamma_{\alpha\beta} \de t^\alpha \de \bar t^\beta
+  e^{\tfrac12(\rho+R)} \tilde\gamma^{\alpha\beta} (\de w_\alpha +\mathcal{W}_\alpha)
(\de \bar w_\beta +\bar{\mathcal{W}}_\alpha).
\ee
The exponential factor in front of the first term corresponds to the inverse volume of the elliptic
fiber $\cE$. In the limit $R\to-\infty$, the volume of the base $\cB$ is given by
$e^{-\tfrac12(\rho+R)}$ (the inverse of the exponential factor in front
of the second term) and is much larger than the fiber volume. When this happens,
the metrics $\gamma_{\alpha\beta}$, $\tilde\gamma^{\alpha\beta}$ and the connection
$\mathcal{W}_{\alpha}$ are expected to depend only on the spectral curve moduli $t^\alpha$,
so that the metric is flat along the torus fibers.
In this case, both $\tilde\gamma^{\alpha\beta}$ and $\mathcal{W}_{\alpha}$ are fixed
from the \kahler metric $\gamma_{\alpha\beta}$ by requiring the total space to be \hk
(see \cite{zbMATH01582122} for a construction of the semi-flat \hk metric on the cotangent
bundle of a \kahler manifold).

\subsection{Kalb-Ramond moduli \label{sec_kr}}

Finally, we consider the $B$-field moduli. For given fixed metric and gauge bundle,
they parametrize solutions of the field
equations
\be
\de \star_g \de B + \frac{1}{4}\, \de \star_g ( \omega_{G} - \omega_{L} ) = 0.
\ee

To compute the metric on the $B$-moduli space,  it is convenient to first dualize the ten-dimensional
Kalb-Ramond two-form $B$ into a 6-form potential $C_6$, which leads to the following action
\be
S_{10} = \int_{\IR^6\times \Kf} \left[ -\frac12\, \de C_6\wedge \star \de C_6
+\frac{1}{4}\,\de C_6 \wedge\left( \omega_{G} - \omega_{L} \right) \right] .
\ee
The Kaluza-Klein reduction on $\Kf$ proceeds by decomposing $C_6  = \omega_I\wedge C_4^{I}$
where  $C^I_4$ are 4-forms in the six non-compact dimensions (indices are raised and lowered
using $\eta_{IJ}$). In the six-dimensional Einstein frame, the resulting action reads
\be
S_6 = \int_{\IR^6}  \left[ -\frac12\, e^{-\rho}\, M_{IJ}\, \de C^I_4 \wedge \star \de C^J_4
+ \frac{1}{4}\, \de C^I_4\wedge  \int_{\Kf} \omega_I \wedge
\left( \omega_{G} - \omega_{L} \right)  \right] ,
\ee
where we used \eqref{defMIJ}.
The integral in the last term turns the 5-form $\omega_I \wedge
( \omega_{G} - \omega_{L})$ on $\IR^6 \times \Kf$ into a one-form $\cV_I$ on $\IR^6$.
Dualizing the 4-forms $C^I_4$ into compact scalars $b_I$ in 6 dimensions leads finally to a metric
\be
\label{dsB}
e^{\rho} \,\de s^2_{B} = e^{\rho} \, M^{IJ}\, (\de b_I + \cV_I ) (\de b_J + \cV_J).
\ee
It describes a torus bundle of rank $n+2$,
with constant metric along the torus action but non-trivial curvature over the space of metric and bundle moduli.
Denoting this curvature by $c_1(b_I)$, one finds
\be
\label{c1B}
c_1(b_I) = \de \cV_I =  \frac{1}{4}
\int_{\Kf} \omega_I \wedge ( \Tr F\wedge F - \Tr R\wedge R )\,  .
\ee
The fact that
the coordinates $b^I$ are compact with period one requires that the
curvatures $c_1(b_I)$ should have integer periods on any
non-contractible two-cycle inside the moduli space $\cM_{g,F}$ over
which the torus $T^{n+2}$ parametrized by $b^I$ is fibered.

This description was used in \cite{Witten:1999fq} to argue that the hypermultiplet moduli space near
an $A_1$ singularity, in the limit where gravity decouples,  is the moduli space of $SU(2)$,
charge one instantons, also known as the Atiyah-Hitchin manifold. The flatness of
the metric along $T^{n+2}$ holds only in the large volume limit, since worldsheet
instantons wrapping genus zero curves in $\Kf$ depend non-trivially on the $B$-moduli.
The metric $M_{IJ}$ may as well receive perturbative $\alpha'$ corrections at finite volume.
The connection $\cV_I$, however, is fixed by the Chern-Simons coupling in 10 dimensions
and related to the  chiral anomaly on the heterotic worldsheet \cite{Hull:1986xn}, so  is
expected to be exact.

Let us now extract the component of the curvature \eqref{c1B} along the bundle moduli space $\cM_F(g)$.
The relevant part of the field strength $F$ that contributes to \eqref{c1B} is $\pa_x \cA+{\rm c.c.}$,
where the derivative is along the non-compact directions, leading to
\begin{eqnarray}
\label{c1Bbndlmd}
\left.c_1(b_I)\right|_{\cM_F(g)} &=&
-2  \cM_{I,\bar k{\ell}} \,\d\bar\xi^{\bar k} \wedge \d{\xi}^{\ell}
- \left( \cN_{I,k\ell} \,\d\xi^k\wedge \d\xi^\ell \,+\, c.c. \right)  ,
\end{eqnarray}
where
\be
\label{defcMN}
\cM_{I,{\bar k}\ell} =  \frac{1}{4}
\int_{\Kf} \omega_I \wedge \Tr \,D_{\bar{\xi}^{\bar k}}\cA^\dagger\wedge
D_{\xi^\ell}\cA\, ,
\qquad
\cN_{I,k\ell} =  \frac{1}{4} \int_{\Kf} \omega_I \wedge \Tr \,D_{\xi^k}\cA\wedge D_{\xi^\ell}\cA\, .
\ee
Notice that $\Tr D_{\xi^k}\cA\wedge D_{\xi^\ell}\cA$ and
$ \Tr \,D_{\bar{\xi}^{\bar k}}\cA^\dagger\wedge  D_{\xi^\ell}\cA$
 are well defined standard (closed) two-forms
(i.e. they are singlets with respect to the gauge group), respectively of degree $(0,2)$ and $(1,1)$. The former
will then be proportional to the only $(0,2)$ form, i.e. the complex conjugate of the holomorphic two-form $\cJ$, while
the latter will have a component along the K\"ahler form $\cJ^3$ and one piece orthogonal to it. In particular
\be
\begin{split}
\Tr \,D_{\bar{\xi}^{\bar k}}\cA^\dagger\wedge  D_{\xi^\ell}\cA =&\,
\left(\frac{\int_\Kf \cJ^3\wedge \Tr \,D_{\bar{\xi}^{\bar k}}\cA^\dagger\wedge
D_{\xi^\ell}\cA}{\int_\Kf \cJ^3\wedge \cJ^3} \right) \cJ^3
+ \left. \Tr \,D_{\bar{\xi}^{\bar k}}\cA^\dagger\wedge  D_{\xi^\ell}\cA\right|_\perp
\\
=&\, \frac{\I}{2}\, e^\rho\, \cG_{{\bar k}{\ell}}  \cJ^3
+ \left. \Tr \,D_{\bar{\xi}^{\bar k}}\cA^\dagger\wedge  D_{\xi^\ell}\cA\right|_\perp\, ,
\end{split}
\ee
where $\left. \Tr \,D_{\bar{\xi}^{\bar k}}\cA^\dagger\wedge  D_{\xi^\ell}\cA\right|_\perp$ is the component orthogonal to $\cJ^3$.
In the second line we have used the fact that Hodge duality on a one-form $\alpha$ on $\Kf$ acts as
$\star\, \alpha = - \I \,  \cJ^3 \wedge \alpha$, so that
\be
\int_{\Kf} \cJ^3 \wedge \Tr \,D_{\bar{\xi}^{\bar{k}}}\cA^\dagger\wedge D_{\xi^\ell}\cA
=  \I  \int_{\Kf} \Tr \(\star \,D_{\bar{\xi}^{\bar{k}}}\cA^\dagger\)\wedge D_{\xi^\ell}\cA = \I \, \cG_{{\bar k}{\ell}} \:.
\ee
These relations allow to write the coefficients $\cM_{I,\bar{k}\ell}$ and $\cN_{I,k\ell}$ as
\begin{eqnarray}
\cM_{I,\bar{k}\ell} &=&  \frac{\I}{8} \,e^\rho\,  \cG_{{\bar k}{\ell}}
\int_\Kf \omega_I \wedge \cJ^3 + \widetilde{\cM}_{I,{\bar k}{\ell}}
\qquad \mbox{with}\qquad
\widetilde{\cM}_{I,\bar{k}\ell} \,\gamma^I_x =0\,,
\label{matrixMIkl}\\
\cN_{I,k\ell} &=& \frac{1}{4}\,e_{k\ell} \int_\Kf \omega_I \wedge \bar{\cJ} =\frac{1}{4}\, e^{-\rho/2}\bar{\gamma}_I e_{k\ell} \:.
\label{resN}
\end{eqnarray}
In particular, the results \eqref{matrixMIkl} imply that the curvatures $ \cM^I_{\bar{k}\ell}$ satisfy
\be
\label{Mprop}
\gamma_I \cM^I_{\bar{k}\ell}=0,
\qquad
\gamma^3_I\cM^I_{\bar{k}\ell}= 
\frac{\I}{4} \, e^{\rho/2}\, \cG_{\bar{k}\ell}.
\ee

\subsection{Two-stage fibration and standard embedding locus}

The upshot of this discussion is that in the  limit where the volume $\cV=e^{-\rho}$ of $\Kf$ is large,
the metric on the hypermultiplet moduli space $\cM_{\rm H}$ takes the form
\be
\de s^2 =\de s^2_{g} + e^{\rho/2} \de s^2_F + e^{\rho} \de s^2_{B}
\label{fullmetmat}
\ee
exhibiting a two-stage fibration structure
\be
\label{doublefib2}
\begin{array}{rcc}
 T^{n+2}& \longrightarrow \qquad \cM_{\rm H} \qquad\qquad &
 \\
&  \downarrow &
\\
& \cM_{F}(g)\ \longrightarrow \ \cM_{g,F}\qquad \qquad\qquad &
\\
&  \downarrow &
\\
&  \IR^+\times 
\frac{SO(3,n-1)}{SO(3)\times SO(n-1)}
\end{array}
\ee
Here, $\cM_{F}(g)$ is a \hk space parametrizing anti-self dual connections (of fixed topological type)
on $\Kf$ with fixed metric $g$, of quaternionic dimension $m$ given in \eqref{mdim1}, 
while the torus $T^{n+2}$ parametrizes the $B$-field. 
The metric along the bundle moduli $\cM_{F}(g)$ scales like $e^{\rho/2}$
and shrinks in the large volume limit $\rho\to -\infty$, while the metric along
the $B$-field moduli scales like $e^{\rho}$ and is even smaller. When $\Kf$ is elliptically
fibered, with a base much larger than the fiber, the moduli space $\cM_{F}(g)$ acquires the
structure of a complex integrable model, with a semi-flat \hk metric. The torus $T^{n+2}$
is non-trivially fibered over the moduli space $\cM_{g,F}$ of metrics and bundles, 
with curvature given by \eqref{c1B}, while the bundle moduli space $\cM_{F}(g)$ 
is non-trivially fibered over the metric moduli space.

It is important to note that the metric \eqref{fullmetmat}
was obtained by reducing tree-level supergravity in 10 dimensions,
which is not by itself supersymmetric. Thus, it is not expected to be QK. To obtain a metric
consistent with supersymmetry, one should
perform the Kaluza-Klein reduction of the fully supersymmetric, $R^2$-corrected supergravity
action \cite{Bergshoeff:1989de}, a daunting task that we leave for future work.

A special class of anti-self dual connections is provided by deformations of the spin connection for the HK metric
on $\Kf$. This leads to bundles with $c_2(F)$=24 and structure group $SU(2)$, whose commutant inside
$E_8\times E_8$ is $E_7\times E_8$. The bundle moduli space has dimension $m=45$, leading
to a $65$-dimensional  hypermultiplet space $\cM_{\rm H}$ of the type above, with $n=20$.
Inside this space, there  exists a submanifold corresponding to the locus where the
gauge connection is equal (up to gauge transformation) to the spin connection. Since the
worldsheet SCFT has enhanced $(4,4)$ supersymmetry at this point, its moduli space is entirely
determined to be the symmetric space $SO(4,20)/SO(4)\times SO(20)$ \cite{Seiberg:1988pf}.
This requires a delicate cancellation between the connection terms appearing in the metric on $\cM_F(g)$,
which will remain after freezing the bundle moduli, against perturbative corrections in the
sigma model.

\section{Two-stage fibration from heterotic/type II duality \label{sec_hetcmap}}

In this section, we use heterotic/type II duality to get insight into the structure of the two-stage
fibration \eqref{doublefib2}, in particular into the fibration of the Kalb-Ramond torus $T^{n+2}$
over the bundle and metric moduli. The key idea is that the large volume limit $\rho\to-\infty$ on
the heterotic side (combined with $R\to -\infty$ with $|R|\ll |\rho|$) corresponds to weak coupling on the type IIB side.
In this limit, the hypermultiplet metric is obtained by the $c$-map procedure \cite{Ferrara:1989ik}
from the moduli space of \kahler structure deformations. Our aim is to express the $c$-map metric in terms of heterotic variables,
expose the  fibration structure and read off the corresponding connections.

\subsection{Heterotic/type II duality in hypermultiplet sector}

We first recall the basic features of heterotic/type II duality
in the hypermultiplet sector
\cite{Aspinwall:1998bw,Aspinwall:1999xs,Aspinwall:2000fd,Louis:2011aa,Alexandrov:2012pr}.\footnote{Since
the hypermultiplet moduli space is independent of the $T^2$ moduli,
one may as well work in 6 dimensions, and use heterotic/F-theory duality, where F-theory
is compactified on the same CY threefold $\CYm$.}
According to this duality, $E_8\times E_8$ heterotic string theory
compactified on $\Kf\times T^2$ is equivalent to type IIA string theory compactified on a
Calabi-Yau threefold $\CY$, or to  type IIB string theory compactified on the mirror
CY threefold $\CYm$. The topology of $\CY$ and $\CYm$ depends on the topology of the gauge bundle
on the heterotic side, but a general fact is that both $\CY$ and $\CYm$ must
admit a $K3$ fibration \cite{Klemm:1995tj,Aspinwall:1995vk,Louis:2011aa}.
We shall focus on the type IIB description and denote by $\Sigma$ the fiber
in the $K3$ fibration $\Sigma \rightarrow \CYm \rightarrow \CP$.

In general, the HM moduli space in type IIB string theory compactified on
$\CYm$ is parametrized by the four-dimensional
string coupling $g_{(4)}\equiv 1/\sqrt{r}$, the \kahler moduli
$z^a$, the periods $\zeta^\Lambda, \tzeta_\Lambda$ of the Ramond-Ramond potentials on
$H_{\rm even}(\CYm,\IZ)$ and the Neveu-Schwarz axion $\sigma$, dual to the Kalb-Ramond two-form in 4 dimensions.
The duality identifies \cite{Louis:2011aa,Alexandrov:2012pr}
\be
\label{maprs}
r=\frac12\, e^{-\tfrac12(\rho+R)}\, ,
\qquad 
{\Re s} = e^{-\tfrac12(\rho-R)}\, ,
\ee
where ${\Re s}$ is the area of the base $\CP$ of the $K3$ fibration, so the large volume limit
$\rho\to-\infty $ on the heterotic side corresponds to $g_{(4)}\to 0$ and $\Re s\to +\infty$
on the type IIB side. The ten-dimensional string coupling is however finite in this
 limit \cite{Alexandrov:2012pr},
\be
\frac{1}{g_s} = 4\sqrt{r}\, e^{\cK/2}=
\frac{|X^0|\, e^{-R/2}}{\sqrt{\eta^{AB} X_A\bX_B}} \, .
\ee
Thus, in order for all quantum corrections on the type IIB side to be exponentially suppressed,
one should further take $R\to -\infty$, with $|R|\ll |\rho|$ so that that ${\Re s}$ remains very large.

In this limit, the hypermultiplet moduli space on the type IIB
side is obtained by the (local) $c$-map procedure
from the \kahler moduli space of $\CYm$ \cite{Cecotti:1989qn,Ferrara:1989ik}. In the limit
$\Re s\to +\infty$, the prepotential takes the form
\be
\label{FclasG}
F(Z^\Lambda)=-\frac{Z^s\eta_{ij}Z^i Z^j}{2Z^0} + f(Z^0,Z^i,Z^\alpha) +  \cO(e^{2\pi\I Z^s/Z^0}) \, ,
\ee
where $Z^s/Z^0=\I s$ is the complex \kahler modulus associated to the base while $Z^i/Z^0=\I t^i$
are complex \kahler moduli for two-cycles $\gamma^i$ in the $K3$ fiber $\Sigma$, with $\eta_{ij}$ being their intersection matrix.
The remaining variables $Z^\alpha/Z^0=\I t^\alpha$ are complex \kahler moduli for the remaining two-cycles $\gamma^\alpha$
in $H^2(\CYm,\IZ)$, dual to reducible singular fibers of the $K3$ fibration \cite{Louis:2011aa}. These
singular fibers do not intersect the section, hence the corresponding \kahler moduli $t^\alpha$
do not appear in the leading term in \eqref{FclasG}.
In contrast, the next-to-leading order term $f(Z^0,Z^i,Z^\alpha)=
(Z^0)^2 f(1,\I t^i, \I t^\alpha)$ does depend on all \kahler moduli except $s$, and includes effects
of all worldsheet instantons wrapping the two-cycles $\gamma^i, \gamma^\alpha$, in addition to the classical cubic contribution,
because the \kahler moduli $t^i, t^\alpha$ stay finite in the large volume limit $\rho\to-\infty$ on the heterotic side.

In \cite{Louis:2011aa,Alexandrov:2012pr}, it was shown that
in the absence of reducible singular fibers, the QK metric derived from the leading term
\eqref{FclasG} describes the symmetric space $SO(4,n)/SO(4)\times SO(n)$, reproducing
the expected hypermultiplet moduli space for heterotic strings compactified on $K3$ equipped
with a rigid gauge bundle. Under this identification, the heterotic and type II variables were
related by \eqref{maprs} supplemented by
\be
\Re(\I s)=B_2,
\qquad
c^A= v^A ,
\qquad
\tc_A=B_A-B_2 v_A ,
\qquad
\sigma=- 2B_1 - v^A\, B_A ,
\label{mirmaptree}
\ee
where $c^A,\tc_A$ are related to $\zeta^A,\tzeta_A$ by a symplectic rotation
$(Z^s,F_s)\mapsto (F_s,-Z^s)$, whereas the moduli $t^i$ are related to the complex structure moduli $X^A$
on the heterotic side via \eqref{newbasegen} below.
Here the index $A$ runs over the $n$ values $(0,s,i)$ so that $\Lambda=(A,\alpha)$.
The additional moduli $t^\alpha$ and $c^\alpha,\tc_\alpha$
associated with reducible bad
fibers can be identified with bundle moduli on the heterotic side \cite{Aspinwall:1998bw,Aspinwall:2000fd}.
The leading contribution to the metric along these directions comes from the second term in \eqref{FclasG},
which can no longer be ignored. As we shall now demonstrate, the QK space obtained by the $c$-map procedure
from the prepotential \eqref{FclasG}, keeping only the first two terms\footnote{All other
corrections to the prepotential correspond to the worldsheet instantons wrapped on the base $\CP$ of
the $K3$ fibration and are exponentially suppressed in our scaling limit.
Thus, $f$ is the only relevant correction.},
has the two-stage fibration
structure \eqref{doublefib2}. Furthermore, the metric on the HK fiber $\cM_F(g)$ is the rigid $c$-map space
\cite{Cecotti:1989qn} derived from the prepotential $f(Z^0,Z^i,Z^\alpha)$ for fixed $Z^0, Z^i$.

\subsection{$c$-map in heterotic variables \label{sec_dualvar}}

The QK metric obtained by the  $c$-map procedure from a special \kahler manifold with
holomorphic symplectic section $\Omega(z^a)=(Z^\Lambda,F_\Lambda)$ takes
the form
\be
\begin{split}
\de s^2=& \, \frac{1}{r^2}\,\de r^2
-\frac{1 }{2r}\, (\Im\cN)^{\Lambda\Sigma}\(\de\tzeta_\Lambda -\cN_{\Lambda\Lambda'}\de\zeta^{\Lambda'}\)
\(\de\tzeta_\Sigma -\bar\cN_{\Sigma\Sigma'}\de\zeta^{\Sigma'}\)
\\ &
+ \frac{1}{16r^2}\(\de\sigma+\tzeta_\Lambda\de\zeta^\Lambda-\zeta^\Lambda\de\tzeta_\Lambda\)^2
+4\cK_{a\bar b}\,\de z^a \de \bz^{\bar b}  ,
\end{split}
\label{hypmettree}
\ee
where $\cK_{a\bar b}$ is the \kahler metric on the  special K\"ahler manifold,
\be
\cK_{a\bar b} = \pa_{z^a}\pa_{\bar z^b} \cK,
\qquad
\cK=-\log\left[
\I \left(\bZ^\Lambda F_\Lambda - Z^\Lambda \bF_\Lambda\right)\right],
\label{Kahlerpot_spec}
\ee
and  $\cN_{\Lambda\Sigma}$ is a symmetric complex matrix with negative definite imaginary part,
determined by the conditions \cite{Ceresole:1995jg,Ceresole:1995ca}
\be
F_\Lambda = \cN_{\Lambda\Sigma} Z^\Sigma\, ,
\qquad
D_a F_\Lambda = \bar\cN_{\Lambda\Sigma} D_a Z^\Sigma\, ,
\ee
where $D_a=\pa_a +\pa_a \cK$ is the \kahler covariant derivative.
When $\Omega(z^a)$ derives from a homogeneous prepotential $F(Z^\Lambda)$, i.e.
$F_\Lambda\equiv \pa_{Z^\Lambda} F$, the matrix
$ \cN_{\Lambda\Sigma} $
is given in terms of the second derivative $F_{\Lambda\Sigma}=\pa_{Z^\Lambda}\pa_{Z^\Sigma}
F$ via
\be
\cN_{\Lambda\Sigma} =\bF_{\Lambda\Sigma} +
2\I\, \frac{ \Im F_{\Lambda\Lambda'}Z^{\Lambda'} \Im F_{\Sigma\Sigma'}Z^{\Sigma'}}{Z^\Xi \Im F_{\Xi\Xi'}Z^{\Xi'}}\, .
\ee

However, for the purposes of heterotic/type II duality, it is more convenient to work in a different
symplectic basis where the prepotential does not exist \cite{Ceresole:1995jg,deWit:1995zg}.
The section $(X^\Lambda,G_\Lambda)$
in this new basis is related to the section $(Z^\Lambda,F_\Lambda)$ in which \eqref{FclasG} applies
by the symplectic transformation on the symplectic plane associated to the \kahler modulus
$s$
\be
X^s=F_s,
\qquad
G_s=-Z^s.
\ee
We denote by $c^\Lambda,\tilde c_\Lambda$ the coordinates
$\zeta^\Lambda,\tilde\zeta_\Lambda$ in this new basis.
The new symplectic section is given by
\be
X^\Lambda=X^0(1,\haf\,\eta_{ij}t^i t^j,\I t^i,\I t^\alpha),
\qquad
G_\Lambda=-\I s\, X_\Lambda +f_\Lambda,
\label{newbasegen}
\ee
where $f_\Lambda=(\pa_{Z^0},0,\pa_{Z^i},\pa_{Z^\alpha})f$ and $X_\Lambda=\eta_{\Lambda\Sigma} X^\Sigma$ with
$\eta_{\Lambda\Sigma}$   the degenerate symmetric matrix
\be
\label{defetaLS}
\eta_{\Lambda\Sigma}=\( \begin{array}{cc}
\eta_{AB} & 0
\\
 0 & 0
\end{array}\),
\qquad
\eta_{AB}=\(\begin{array}{ccc}
0 & 1 & 0
\\
1 & 0 & 0
\\
0 & 0 & \eta_{ij}
\end{array}\).
\ee
Note that $X^A$ satisfy the same constraint $X^A\eta_{AB} X^B=0$ as in section \ref{secmetmod} upon identification of $\eta_{AB}$
with the intersection form of 2-cycles on $K3$.

The \kahler potential and period matrix read
\be
\begin{split}
\cK=& -\log \(\Re s-\matf \)-\log|t+\bt|^2-\log |X^0|^2,
\\
\cN_{\Lambda\Sigma}=&
\,\I \bs\, \eta_{\Lambda\Sigma}-\I(s+\bs)X_{\Lambda\Sigma}
+\bar f_{\Lambda\Sigma}\\
& +2\I\,\frac{\bX_\Lambda X^\Xi\Im f_{\Sigma\Xi}+\bX_\Sigma X^\Xi\Im f_{\Lambda\Xi}}{\eta_{CD}X^C\bX^D}
-2\I\, \frac{\bX_\Lambda\bX_\Sigma}{\(\eta_{CD}X^C\bX^D\)^2}\, X^\Xi X^\Theta \Im f_{\Xi\Theta},
\end{split}
\label{newNgen}
\ee
where $|t+\bt|^2=\eta_{ij}(t^i+\bt^i)(t^j+\bt^j)$ and we introduced
\be
\label{defmatf}
\matf=\hf\, X^{\Lambda\Sigma}\Im f_{\Lambda\Sigma}\ ,\qquad
X^{\Lambda\Sigma}=\frac{X^\Lambda\bX^\Sigma+\bX^\Lambda X^\Sigma}{\eta_{CD}X^C\bX^D}\, ,
\ee
while indices on $X^{\Lambda\Sigma}$ are lowered using \eqref{defetaLS}.
The function $\matf$, a real function of $t^i,t^\alpha$, will play an important
role below. It determines the order $\cO(1/\Re s)$ correction to the \kahler potential 
$\cK$ in the limit $\Re s\to +\infty$. The matrix $X^{\Lambda\Sigma}$ generalizes $X^{AB}$ in \eqref{matrixMAB}.
Splitting the index $\Lambda$ into $A,\alpha$, the inverse of the imaginary part
of the matrix $\cN_{\Lambda\Sigma}$ can be computed to be
\beq
\Im\cN^{AB}&=&
\(\delta^A_{C}-{X^A}_C\)V^{CD} \(\delta^B_{D}-{X_D}^B\)
-\frac{X^{AB}}{\Re s-\matf},
\nn\\
&&
\nn\\
\Im\cN^{A\alpha}&=&-\(\delta^A_{C}-{X^A}_C\)V^{CD} \(\delta^B_{D}-{X_D}^B\)\nu_{B\beta}\mu^{\alpha\beta}
-\frac{X^{A\alpha}}{\Re s-\matf},
\label{inverseImNgen}\\
\Im\cN^{\alpha\beta}&=&
-\mu^{\alpha\beta}
+\mu^{\alpha\gamma}\nu_{A\gamma}\(\delta^A_{C}-{X^A}_C\)V^{CD} \(\delta^B_{D}-{X_D}^B\)\nu_{B\delta}\mu^{\delta\beta}
-\frac{X^{\alpha\beta}}{\Re s-\matf},
\nn
\eeq
where we introduced the following notations:
\be
\begin{split}
\label{matXMgen}
\mu_{\alpha\beta}=&\,\Im f_{\alpha\beta},
\qquad
\nu_{A\alpha}=\Im f_{A\alpha},
\qquad
\lambda_{AB}=\Im f_{AB},
\\
V_{AB}=&\,\Re s\,\eta_{AB}-\(\delta_A^{C}-{X_A}^C\)\(\lambda_{CD}-\nu_{C\alpha}\mu^{\alpha\beta} \nu_{D\beta}\) \(\delta_B^{D}-{X^D}_B\),
\end{split}
\ee
and $\mu^{\alpha\beta}$, $V^{AB}$ are the inverse of $\mu_{\alpha\beta}$ and $V_{AB}$, respectively.

We now apply the same duality map \eqref{mirmaptree} and \eqref{maprs}, except for a minor change
in the definition of the coordinates $\rho,R$ involving the function  $\matf$,
\be
\label{maprs2}
r=\frac12 \,e^{-\tfrac12(\rho+R)},
\qquad
{\Re s} = e^{-\tfrac12(\rho-R)} + \matf ,
\ee
\be\qquad
\Re(\I s)=B_2 ,
\qquad
c^A= v^A,
\qquad
\tc_A=B_A-B_2 v_A,
\qquad
\sigma=- 2B_1 - v^A\, B_A.
\label{mirmaptree2}
\ee
Guided by similar definitions in the absence of bundle moduli $t^\alpha$
\cite{Louis:2011aa,Alexandrov:2012pr}, we further define the symmetric matrix
\be
\cM_{AB}=M_{AB}+e^{\hf(\rho-R)}\(\delta_A^{C}-{X_A}^C\)\(\lambda_{CD}-\nu_{C\alpha}\mu^{\alpha\beta} \nu_{D\beta}-\matf\, \eta_{CD}\)
\(\delta_B^{D}-{X^D}_B\)
\ee
with $M_{AB}$ from \eqref{matrixMAB},
in such a way that its inverse $\cM^{AB}$ satisfies $\cM^{AB}=-e^{\hf(R-\rho)}\Im \cN^{AB}$.
$\cM_{AB}$ reduces to  $M_{AB}$ in the limit $\rho\to-\infty$, but in general is {\it not} an element of $SO(2,n-2)$. We also
introduce a symmetric matrix
\be
\cM^{IJ}=
\(\begin{array}{ccc}
e^R & -\frac{v^2}{2}\, e^R & e^R v^B
\\
-\frac{v^2}{2}\, e^R & \frac{v^4}{4}\, e^R+ e^{-R}+\cM^{AB}v_Av_B & -\frac{v^2}{2}\,e^R v^B-\cM^{BC}v_C
\\
e^R v^A & -\frac{v^2}{2}\,e^R v^A-\cM^{AC}v_C & e^R v^A v^B + \cM^{AB}
\end{array}\)
\label{newbigmat}
\ee
which reduces to the inverse of the matrix $M^{IJ}$ from \eqref{defMIJ} in the limit $\rho\to-\infty$, but
 is in general not an element of $SO(3,n-1)$, except when $\cM^{AB}$ is
an element of $SO(2,n-2)$.

With these definitions, the $c$-map metric \eqref{hypmettree} associated to the
prepotential \eqref{FclasG} (retaining only the first two terms, and
assuming no special property of $f(Z^0,Z^i,Z^\alpha)$ other than independence on $Z^s$ and homogeneity)
can be written as a sum of three contributions,
\be
\de s^2 =\de s^2_{g} + e^{\rho/2} \de s^2_F + e^{\rho} \de s^2_{B} \, ,
\label{fullmetmat2}
\ee
matching the expected form \eqref{fullmetmat} of the hypermultiplet metric moduli space
on the heterotic side.
In the following we discuss each contribution in turn.

\subsubsection{Metric moduli}

The first term
\be
\label{dsgB}
\de s^2_g =  \hf\, \de \rho^2+
\hf\, \de R^2 -\frac14\, \de M_{AB}\de M^{AB}+e^R \cM_{AB} \de v^A \de v^B
\ee
generalizes the $SO(3,n-1)$
invariant metric \eqref{dsg} on the moduli space $\cM_g$ of HK metrics on $\Kf$.
It reduces to this invariant metric in the large volume limit $\rho\to-\infty$,
but contains in addition power-suppressed corrections
due to the difference between $\cM_{AB}$ and $M_{AB}$.

\subsubsection{Bundle moduli}

The second term is given by
\be
\label{dsFB}
\de s^2_{F}=
4 e^{-R/2}\p\bar\p\matf
+e^{R/2} \mu_{\alpha\beta} Dc^\alpha \, Dc^\beta
+e^{R/2}\mu^{\alpha\beta} (D\tc_\alpha-\Re f_{\alpha\alpha'} \de c^{\alpha'})
(D \tc_\beta-\Re f_{\beta\beta'} \de c^{\beta'}),
\ee
where
\be
\p\bar\p\matf=
\frac{\mu_{\alpha\beta}D X^\alpha D\bX^\beta}{\eta_{CD} X^C\bX^D}
+\(\delta_A^{C}-{X_A}^C\)
\(\lambda_{CD}-\nu_{C\alpha}\mu^{\alpha\beta} \nu_{D\beta}-\matf\, \eta_{CD}\)
 \(\delta_B^{D}-{X^D}_B\) \frac{\de X^A \de \bX^B}{\eta_{CD} X^C\bX^D} .
\label{ddfterm}
\ee
It contains kinetic terms for the bundle
moduli\footnote{It also contains a contribution to the kinetic term for the metric moduli $t^i$
generated by the second term in \eqref{ddfterm}, which is however suppressed with respect to the leading
contribution from $\de s^2_g$.}
$t^\alpha,c^\alpha$ and $\tc_\alpha$, with specific connections with respect to the metric moduli,
\be
D X^\alpha=\de X^\alpha-\Gamma_A^\alpha\de X^A,
\qquad
Dc^\alpha =\de c^\alpha - \Gamma^{\alpha}_A \de v^A,
\qquad
D\tilde c_\alpha =  \de \tilde c_\alpha -\Re\cN_{\alpha A}\de v^{A},
\ee
where
\be
\Gamma_A^\alpha=\mu^{\alpha\beta}\Im \cN_{A\beta}={X_A}^\alpha-\(\delta_A^{B}-{X_A}^B\)\nu_{B\beta}\mu^{\alpha\beta}.
\ee
Setting these connection terms to zero, i.e. fixing the HK metric on $\Kf$, the metric \eqref{dsFB} reduces to
\be
\label{dsFB2}
\left.\de s^2_{F}\right\vert_{g} =
8\,e^{-R/2} \,\frac{\mu_{\alpha\beta}\de t^\alpha \de \bar t^\beta}{|t+\bt|^2}
+ e^{R/2}\mu^{\alpha\beta} \, (\de w_\alpha + \cW_\alpha)\, (\de \bar w_\beta + \bar\cW_\beta),
\ee
where we denoted
\be
w_\alpha=\tilde c_\alpha- f_{\alpha\beta} c^\beta,
\qquad
\cW_{\alpha} =- \frac{1}{2}\, \mu^{\gamma\lambda}(w_\lambda-\bw_\lambda)f_{\alpha\beta\gamma}\de t^\beta.
\ee
We recognize in this expression
the rigid $c$-map metric \cite{Cecotti:1989qn} associated to the prepotential $f(Z^0,Z^i,Z^\alpha)$
for fixed $Z^0,Z^i$ with $(t^\alpha,w_\alpha)$ being holomorphic Darboux coordinates
for the holomorphic symplectic form \eqref{holsymF} on $\cM_F(g)$.
Eq. \eqref{dsFB2} is in agreement with the expected form of the metric \eqref{dsFcover} in coordinates
adapted to the spectral cover construction. In particular, the real function $\matf$
defined in \eqref{defmatf} provides, up to an overall factor $4e^{-R/2}$, a \kahler potential
for the \kahler metric on the moduli space $\IP^g$ of spectral covers mentioned above \eqref{dsFcover}.
It is somewhat unexpected that the metric \eqref{dsFB} on bundle moduli space should
belong to the class of rigid $c$-map metrics, rather than the more general class
of semi-flat HK metrics on cotangent bundles of \kahler manifolds constructed in \cite{zbMATH01582122}.

\subsubsection{Kalb-Ramond moduli}

Finally,
\be
\label{dsBB}
\de s^2_{B} = \cM^{IJ}\, (\de B_I+\cV_I) \, (\de B_J+\cV_J)
\ee
reproduces the flat metric on the torus fiber $T^{n+2}$
of the two-stage bundle \eqref{doublefib2}. The kinetic term $\cM^{IJ}$ is in agreement with
\eqref{dsB} in the large volume limit, while the connection
reads\footnote{The term $B_2\, \eta_{AB}$ in the last term of $\cV_A$ 
cancels the dependence on the $B$-field introduced by $\Re\cN_{AB}$.}
\be
\begin{split}
\cV_1=&\,\frac{v^2}{2}\, \cV_2-v^A\cV_A+\hf\(c^\alpha\de\tc_\alpha-\tc_\alpha\de c^\alpha\),
\\
\cV_2=&\,-\I(\p-\bar\p)\matf,
\\
\cV_A=&-\I v_A(\p-\bar\p)\matf + \,\Gamma_A^\alpha\de\tc_\alpha-\(\Re\cN_{A\beta}+\Gamma_A^\alpha\Re\cN_{\alpha\beta}\) \, \de c^\beta
\\
&\,
-\(\Re\cN_{AB}+B_2\, \eta_{AB}+\Gamma_A^\alpha\Re\cN_{\alpha B}\)\de v^B.
\end{split}
\label{connection-bundle}
\ee
We shall be particularly interested in the restriction of the curvatures $\de \cV_I$
along the bundle moduli directions (i.e. for fixed metric $g$ on $\Kf$), which can be decomposed
into their (1,1) and (2,0) components,
\be
\de \cV_I\vert_g =-2\cM_{I,\bar{k}\ell}\de \bar\xi^{\bar k}\wedge \de\xi^\ell-\(\cN_{I,k\ell}\de \xi^k\wedge\de\xi^\ell+{\rm c.c.}\),
\ee
where $\xi^k=(t^\alpha,w_\alpha)$ denote the holomorphic bundle moduli. Denoting also
\beq
\phi_{A\beta}^\alpha& =&
\frac{1}{4}\(\delta_A^B-{X_A}^B\)\mu^{\alpha\gamma}\( f_{\beta\gamma\lambda}\mu^{\lambda\delta}\Im f_{B\delta}-f_{B\beta\gamma}\),
\\
\psi^\alpha_\beta &=&\frac{\I}{8}\, f_{\beta\gamma\lambda}\mu^{\alpha\gamma}\mu^{\lambda\delta}(w_\delta-\bw_\delta) ,
\eeq
one finds
\be
\begin{split}
\cM_{1,{\bar k}\ell} =&
\(\begin{array}{cc}
 -\frac{\I v^2\mu_{\alpha\beta}}{|t+\bt|^2}
 +4\I v^A\mu_{\gamma\gamma'}\(\bar\phi^\gamma_{A\alpha}\psi_\beta^{\gamma'}+\phi^\gamma_{A\beta}\bar \psi_\alpha^{\gamma'} \)
+4\I \mu_{\gamma\gamma'}\bar \psi_\alpha^\gamma \psi_\beta^{\gamma'}
&
v^A \bar \phi_{A\alpha}^\beta+\bar \psi_\alpha^\beta
\\
-v^A \phi_{A\beta}^\alpha -\psi^\alpha_\beta
& \frac{\I}{4}\, \mu^{\alpha\beta}
\end{array}\),
\\
\cM_{2,\bar{k}\ell} =&
\(\begin{array}{cc}
\frac{2\I\mu_{\alpha\beta}}{|t+\bt|^2} & 0
\\
0 & 0
\end{array}\),
\\
\cM_{A,\bar{k}\ell}=&
\(\begin{array}{cc}
\frac{2\I v_A\mu_{\alpha\beta}}{|t+\bt|^2}
-4\I\mu_{\gamma\gamma'}\(\bar\phi^\gamma_{A\alpha}\psi_\beta^{\gamma'}+\phi^\gamma_{A\beta}\bar \psi_\alpha^{\gamma'} \)
&
-\bar \phi_{A\alpha}^\beta
\\
\phi_{A\beta}^\alpha & 0
\end{array}\)
\end{split}
\ee
and 
\be
\cN_{I,k\ell}
=\frac{\I\bar\gamma_I}{\sqrt{2}|t+\bt|}\(\begin{array}{cc}
0 & -\delta_\alpha^\beta
\\
\delta_\beta^\alpha & 0
\end{array}\).
\ee
In particular, it is easily checked that $\cN_{I,k\ell}$ agrees with \eqref{resN} given that
$t^\alpha$ and $w_\alpha$ are Darboux coordinates for the symplectic matrix $e_{k\ell}$,
whereas $\cM_{I,\bar{k}\ell}$ satisfies \eqref{Mprop} with
\be
\cG_{\bar{k}\ell}=e^{\hf \(R-\rho\)}\(\begin{array}{cc}
\frac{8e^{-R}\mu_{\alpha\beta}}{|t+\bt|^2}+16\mu_{\gamma\gamma'}\bar\psi_{\alpha}^{\gamma}\psi_\beta^{\gamma'} 
& 
-4\I\bar\psi_\alpha^\beta
\\
4\I\psi^\alpha_\beta 
&
\mu^{\alpha\beta}
\end{array}\)
\ee
being (up to an overall factor of $e^{\rho/2}$) the metric on bundle moduli space
read off from \eqref{dsFB2}.
Thus,  heterotic/type II duality predicts the values of the integrals \eqref{defcMN}, in terms
of the holomorphic prepotential $f(t^i,t^\alpha)$ which governs the \hk metric
on bundle moduli space. It would be very interesting to compute the
 integrals \eqref{defcMN} independently and extract the prepotential.

\section{Discussion \label{sec_discuss}}

In this work we have used heterotic/type II duality to shed light on the hypermultiplet
moduli space in heterotic string theory compactified on an elliptically fibered $K3$ surface,
in the limit where the volume is very large and the base is much larger than the fiber.
On the type IIA side, this corresponds to a limit where the  string coupling vanishes
while the size of the base of the $K3$ fibration becomes infinite.
In this limit, the classical cubic term in the prepotential \eqref{FclasG}
dominates the kinetic terms of the metric and B-field moduli,
while the bundle moduli metric is determined by the subleading term in \eqref{FclasG}.
The latter in general contains worldsheet instanton corrections wrapping two-cycles
on the $K3$ fiber and/or singular reducible fibers.
By identifying the \kahler moduli $t^\alpha$ and the corresponding Ramond-Ramond moduli
$c^\alpha, \tilde c_\alpha$ associated to these singular fibers with the bundle moduli $\xi^k$ on the heterotic side,
and using the slightly modified duality map of \cite{Louis:2011aa,Alexandrov:2012pr} for the remaining moduli,
we have found that the resulting moduli space has the two-stage fibration
structure \eqref{doublefib2} expected on the heterotic side.
In particular, the bundle moduli space $\cM_F(g)$
is a torus bundle over a \kahler space equipped with a  semi-flat \hk metric,
as appropriate for a complex integrable system. The HK metric is  obtained by
the rigid $c$-map from the subleading term  $f(Z^0,Z^i,Z^\alpha)$ in the prepotential \eqref{FclasG}
(at fixed $Z^0,Z^i$),  a special case of the class of semi-flat HK metrics on
cotangent bundles of \kahler manifolds constructed in \cite{zbMATH01582122}.
It would be interesting to compare
this prediction with a first principle computation of  the metric on the moduli space of spectral covers.

In addition, the $B$-field moduli
live in a torus $T^{n+2}$ with flat metric along the torus action,
and with non-trivial curvature over both the bundle moduli and the metric moduli.
Remarkably, the component of the curvature along the bundle moduli are determined
by the same holomorphic prepotential $f(Z^0,Z^i,Z^\alpha)$ --- a non-trivial prediction
for the integrals appearing in \eqref{defcMN}.
The metric on the torus $T^{n+2}$, given by the matrix $\cM^{IJ}$ \eqref{newbigmat},
also receives corrections suppressed by inverse powers of the volume.
These corrections are necessary for the metric to be QK, and we expect that they could be derived
by performing a careful Kaluza-Klein reduction of the fully supersymmetric, $R^2$-corrected
heterotic supergravity in ten dimensions \cite{Bergshoeff:1989de}.
Before tackling this daunting task however, a basic problem is to construct a set of gauge-invariant coordinates
on bundle and $B$-field moduli space. While the first issue requires a deeper understanding
of the spectral cover construction, the second problem may be usefully circumvented by
dualizing the Kalb-Ramond field to a six-form potential, as discussed in \S\ref{sec_kr},
and dualize back the resulting four-forms after reduction.

Finally, it is worth reiterating that the above difficulties arise just as
well in the more phenomenologically appealing heterotic compactifications on Calabi-Yau threefolds.
We hope that the insights gained through the present study can stimulate
progress on $\cN=1$ heterotic string vacua.

\medskip

\acknowledgments We are grateful to P. Candelas, R. Donagi, R. Minasian and N. Nekrasov for useful discussions.
S.A., J.L. and R.V. thank the Theory Group at CERN for their kind hospitality
throughout the course of this work.
The work of J.L. was supported by the German Science Foundation (DFG) under the Collaborative
Research Center (SFB) 676 Particles, Strings and the Early Universe.


\providecommand{\href}[2]{#2}\begingroup\raggedright\endgroup

\end{document}